# Reversible Photo-Induced Phase Segregation and Origin of Long Carrier Lifetime in Mixed-Halide Perovskite Films

Subodh K. Gautam[1,a)], Minjin Kim[2], Douglas R. Miquita[1,3], Jean-Eric Bouree[2], Bernard Geffroy[2,4] and Olivier Plantevin[1,b)]

[1] Université Paris-Saclay, CNRS, Laboratoire de Physique des Solides, 91405, Orsay, France.

[2] LPICM, CNRS, Ecole Polytechnique, Institut Polytechnique de Paris, route de Saclay, 91128 Palaiseau, France.

[3] Centro de Microscopia - Universidade Federal de Minas Gerais, Belo Horizonte, MG, 31270-901- Brasil

[4] Université Paris-Saclay, CEA, CNRS, NIMBE, LICSEN, 91191, Gif-sur-Yvette, France

## Abstract

Mixed-halide based hybrid perovskite semiconductors have attracted tremendous attention as a promising candidate for high efficient photovoltaic and light-emitting devices. However, these advanced perovskite materials may undergo phase segregation under light illumination due to halide ion migration, affecting their optoelectronic properties. Here, we report photo-excitation induced phase segregation in triple-cation mixed-halide perovskite films that yields to red-shift in photoluminescence response. We demonstrate that photo-excitation induced halide ion migration leads to the formation of smaller-bandgap iodide-rich and larger-bandgap bromide-rich domains in the perovskite film, where the phase segregation rate is found to follow the excitation power-density as a power law. Results confirm that charge carrier lifetime increases with red-shift in photoluminescence due to the trapping of photo-excited carriers in the segregated smaller-bandgap iodide-rich domains. Interestingly, we found that these photo-induced changes are fully reversible and thermally activated when the excitation power is turned off. A significant difference in activation energies for halide ion migration is observed during phase segregation and recovery process under darkness. Additionally, we have investigated the emission linewidth broadening as a function of temperature which is governed by the exciton-optical phonon coupling. The mechanism of photo-induced phase segregation is interpreted based on exciton-phonon coupling strength in both mixed and demixed (segregated) states of perovskite film.

*Authors for correspondence*: a) subodh-kumar.gautam@csnsm.in2p3.fr (Subodh K. Gautam); b) olivier.plantevin@u-psud,fr (Olivier Plantevin)

## Introduction

Over the past few years, organic-inorganic halide perovskites have attracted tremendous attention as a promising material for low-cost preparation and high efficient photovoltaic and light-emitting devices. These hybrid perovskite materials show unique opto-electronic properties such as high absorption coefficient, long carrier diffusion length and small exciton binding energy [1,2]. To date, various combinations and ratios of monovalent inorganic cations (such as formamidinium ($CH(NH_2)_2^+$ = FA), methylammonium ($CH_3NH_3^+$ = MA), Cs, and Rb) and halide anions ($I^-$ and $Br^-$) have been used for bandgap tuning as well as a means of improving efficiency and stability of perovskite photovoltaic devices [3]. One of the most influential improvement was the incorporation of cesium cations in perovskite structure to fine-tune Goldschmidt tolerance factor towards more structurally stable black phase perovskite with improved performance of photovoltaic devices [3–5]. Therefore, triple-cation based mixed-halide perovskites (Cs, FA, MA)$Pb(I, Br)_3$ have been found more suitable for solar cell application. Unfortunately, even such most advanced perovskite materials still suffer from instabilities of the bandgap when subjected to photo-excitation under continuous light illumination [6,7]. As originally reported by Hoke and coworkers, $MAPb(I_{1-x}Br_x)_3$ undergoes reversible halide phase segregation under light illumination into separate iodide- and bromide-rich domains within the parent phase [6]. In recent years, it is commonly reported that light-induced phase segregation process in mixed halide $MAPb(I_{1-x}Br_x)_3$ and $CsPbBr_xI_{3-x}$ perovskites is initiated by segregation of iodine ions which causes a red-shift in the photoluminescence (PL) [7,8]. On the other hand, phase segregation can also induce a PL blue-shift as reported in $CsPbBr_{1.2}I_{1.8}$ nanocrystals subjected to photo-excitation and external electrical biasing [7]. Furthermore, other reports suggest that variation of energy bandgap of mixed halide perovskite could also be due to lattice distortion as light illumination could generate lattice expansion or contraction [9–11]. Therefore, the existing literature on phase segregation is often subject to controversy.

Another puzzling feature under debate in hybrid perovskites is the unusual long carrier lifetime. Several mechanisms have been proposed to explain the long carrier lifetime of perovskite materials, such as large polarons formation, Rashba effect and photon recycling [12]. Moreover, recombination of excited charge carriers through trap-mediated channels also influences the rate of halide segregation and recombination lifetime [13]. The nature of trap states and trapped carrier type (electrons or holes) in segregated domains are not yet entirely understood although it was reported that electrons are more likely to be trapped than holes [14,15]. Such existing reports imply that excited charge carriers trapping in trap-states and corresponding energy levels within the bandgap could alter the phase segregation dynamics. Recently, it was also demonstrated that excited charge carriers could couple with the perovskite lattice to create a polaron [16–18], which could promote the halide ion segregation. A strong exciton–phonon coupling is expected in such soft and ionic hybrid perovskite materials at room temperature [19,20]. Eventually, the Cs cation-based mixed halide perovskite (partially exchanging FA cation with Cs) unveils lower electron–phonon coupling strength and exhibits high structural stability against photo-induced phase segregation [18]. Therefore, it is interesting to study the role of exciton-phonon coupling in triple-cation mixed halide perovskite (TC-MHP) films during phase segregation process and the role of trapped charge carrier under photo-excitation.

In the present study, we have explored different aspects of photo-excitation induced phase segregation in triple-cation mixed-halide $(MA_{0.17}\,FA_{0.83})_{0.95}Cs_{0.05}\,Pb(I_{0.83}\,Br_{0.17})_3$ perovskites (TC-MHP) films and quantitatively rationalized the spectroscopic processes that occur during phase segregation and recovery processes. We demonstrate the laser power dependent evolution of segregated halide domains and their kinetics with laser exposure time. We performed Time-Correlated Single Photon Counting (TCSPC) measurements along with low-temperature PL studies to read the nature of photo-excited carriers trapping states in phase segregated domains, revealing their role in long decay lifetime. We discuss the emission linewidth broadening with rising temperature, which is governed by exciton-optical phonon coupling in both segregated and remaining non-segregated domains of TC-MHP film. Additionally, recovery kinetics of phase-segregated films is studied under dark at different temperatures and activation energies of halide ion migration are derived.

## Experimental details

The triple-cation mixed halide hybrid perovskite thin films (thickness ~380 nm) were prepared using spin coating technique on glass substrates. At first, for the materials' synthesis, the $(MA_{0.17} FA_{0.83})Pb(I_{0.83} Br_{0.17})_3$ perovskite precursor solution was prepared by dissolving $PbI_2$ (1.1M), FAI (1.0M), $PbBr_2$ (0.2M), MABr (0.2M) in a 44 wt.% mixture of anhydrous N,N-Dimethylformamide (DMF) and Dimethyl sulfoxide (DMSO). The solution was kept on stirring and then 42μL (5 vol.%) of CsI (1.5M in DMSO) was added in the resultant solution at room temperature. The precursors of FAI and $PbI_2$ were purchased from Sigma Aldrich; MABr from Solaronix; $PbI_2$ and CsI from Alfa Aesar. Before thin film coating, glass substrates were initially cleaned in deionized water, acetone, 2-Propanol sequentially by ultra-sonication followed by drying in dry nitrogen gas. In general, a dynamic dispense is preferred in synthesis process as it is a more controlled process that produces less substrate-to-substrate variation. The triple-cation solution was dispensed after the spin revolution was stabilized at 6000 rpm. The total spinning time was 30 s and then 200 μL of chlorobenzene was dropped for 5 s to remove residual solvent. Films were annealed at 100°C in nitrogen ambient for 1 h, allowing solvent evaporation and thin films crystallization.

The surface morphology measurement of prepared TC-MHP films were carried out using field-emission scanning electron microscopy (FE-SEM, JSM-7000 F) at CNRS-CSNSM, Orsay. The UV-visible absorption measurements were performed using Cary 5000 UV-VIS-NIR double beam spectrophotometer. Photoluminescence Spectroscopy studies on mixed halide perovskite sample were performed in reflection geometry with a Horiba-JY Quantamaster spectrometer equipped with a R13456 photomultiplier (Hamamatsu) detector. We used a continuous-wave fiber-coupled laser diode (MDL-III-454nm/90-800mW) from CNI; it had a wavelength of ~ 454±5 nm at different excitation powers and was focused on the sample on a spot size of ~1.75 nm diameter. The time-resolved single photon counting (TCSPC) measurements were carried out using NanoLED source (Horiba), which had an excitation of 482 nm and was pulsed at 25 kHz. For low-temperature measurements the samples were glued with silver paste in an optical closed-cycle cryostat from ARS Instruments equipped with two quartz windows.

## Results and discussion

**Photo-excitation induced phase segregation**

The TC-MHP perovskite film exhibits high optical absorbance and exhibit the direct band gap ($E_g$) of ~ 1.625 eV, calculated using the Tauc's relation as presented in the supplementary Fig. S1. The surface morphological properties of TC-MHP film is also presented in supplementary Fig. S2: a SEM image shows the uniformly distributed nanostructure with average grains sizes of ~175 ± 50 nm. We have performed laser power dependent PL measurements at room temperature (296 K) to provide insight into the photo-excitation induced spectroscopic variation in TC-MHP film in terms of carrier trap-states and emission recombination dynamics. Fig. 1 (a) shows the normalized PL spectra as a function of laser power density ($P_{exc}$). With increasing excitation power density, a continuous redshift of about ~ 58 meV is observed in emission peak: it goes from 1.625 eV (1 mW/cm$^2$) to 1.566 eV (840 mW/cm$^2$), concurrent with the increase in PL emission intensity ($I_{em}$). In Fig. 1 (b), data clearly shows that PL intensity increases as a power law of excitation power, with an exponent ~3/2 (i.e. $I_{em} \propto P_{exc}^b$, with b= 1.48) for power-densities up to 380 mWcm$^{-2}$. Generally, in the case of free excited carrier recombination, $I_{em}$ grows linearly with $P_{exc}$ (i.e., $I_{em} \propto P_{exc}$) [21]. In present case, this possibility is readily exempted at room temperature, because excitonic feature is not observed in the absorption spectrum (supplementary Fig. S1). The unusual ~3/2 power low PL response points the participation of intra-gap states in recombination process that act as traps for electrons or holes. Therefore, superlinear growth of PL intensity suggests that photo-generated carrier trapping at available defects or traps sites competes with bimolecular electron/hole recombination process in TC-MHP film [21,22]. At higher excitation power-density (above 380 mWcm$^{-2}$), the power law changes from superlinear to sublinear owing to the fact that trap-mediated recombination process is dominated by bimolecular recombination process controlled with Auger losses. Therefore, at higher excitation power-density, the saturation in $I_{em}$ is attributed to the saturation of radiative bimolecular recombination centers, as Auger losses starts playing a key role in recombination process.

Fig. 1 (c) shows the red-shift and broadening in PL emission linewidth as a function of excitation power-density which confirms the participation of low energy sub-bandgap states in recombination process. To further confirm the photo-excitation induced red-shift in PL emission, time dependent PL measurements were performed at given excitation power-densities and shown in supplementary Fig. S3 (a-f). At low-excitation power-density (1 mW/cm$^2$), the position of PL spectrum remains unchanged and shows significant increment in PL intensity over exposure time. The increase in PL intensity might be related to the filling of existing non-radiative traps sites in TC-MHP film. At higher excitation power-densities, results show the red-shift and an increase in PL intensity with exposure time, indicating the formation of low-energy sub-bandgap states and filling of existing trap sites [23]. The laser-induced heating possibility is safely ruled out here since temperature dependent PL emission exhibit blue-shift with increasing temperature (discussed in later section). It is observed that increasing excitation power leads to faster shifting of PL emission peak (supplementary Fig. S3). In particular, it is noted that photo-excited TC-MHP film shows red-shifted PL emission at ~1.564 eV corresponds to the standard PL emission from $FAPbI_3$ perovskite. It confirms that the magnitude of redshift is related to the evolution of localized low-bandgap energy states, which might be related to halide ion migration and iodide-rich perovskite phase. Therefore, it establishes the possibility of photo-excitation induced halide ion migration and formation of smaller-bandgap iodide-rich perovskite domains in TC-MHP film that yields red-shift in PL emission.

Fig. 1 (d) shows the redshift in peak position as a function of time at given excitation power densities. Peak position values are extracted using PL emission data (supplementary Fig. S3) fitted with Gaussian-Lorentzian (Voigt) function. The first order phase segregation rate constant ($k_{seg}$) is calculated from subsequent fitting of exponential red-shifting curves at different excitation power densities. It is found that at excitation power ≤ 1 mW/cm$^2$, PL emission peak remains stable at a position of ~1.624 eV and reveals that no phase segregation occurs up to long exposure time. At $P_{exc}$ = 5 mW/cm$^2$, PL peak shifts slowly in exponential manner and shows ~20 meV redshift after 140 min illumination; the corresponding $k_{seg}$ value is found to be 0.21 ×10$^{-3}$ s$^{-1}$. The PL emission peak shifts faster with increasing excitation power-density and exhibits higher $k_{seg}$ value. The maximum redshift in PL emission peak is about ~56 meV for $P_{exc}$ = 84 mW/cm$^2$ and no further shift is observed at higher $P_{exc}$ value. The $k_{seg}$ value increases in a non-linear fashion with increasing $P_{exc}$, from 0.21 ×10$^{-3}$ s$^{-1}$ for $P_{exc}$ = 5 mW/cm$^2$ to 8.92 × 10$^{-3}$ s$^{-1}$ for $P_{exc}$ = 840 mW/cm$^2$. Therefore, high $k_{seg}$ value corresponds to faster exponential growth of iodide-segregated domains. It is interesting to note that $k_{seg}$ in TC-MHP shows lower propensity for halide segregation than the reported MA-based mixed halide perovskite with a similar halide composition range ($0.2 < x < 0.5$) [8,13]. Furthermore, a relationship between PL-redshift and halide composition variation is establish using non-linear bandgap variation as a function of halide composition ratio (Br, x = 0 to 1.0) under Vegard's law expression and presented in Fig. S4 (supplementary information). The PL-emission values of mixed and segregated domains are used to fit the non-linear bandgap variation curve as a function of bromide concentration. Result shows that PL-redshift follows the non-linear decrease in bandgap value with reduction in Br-concentration (enrichment in iodide) in TC-MHP film. It confirms that photo-excitation induces halide composition changes due to halide ion migration and leads to formation of segregated halide domains in TC-MHP film. Moreover, PL emission broadening in mixed TC-MHP may also occur due to the change in halide composition (decrease in bromide and enrichment of iodide content) as interpreted from Fig. 1(c). However, the PL width broadening with varying halide composition may also include other complex phenomenon such as increased lattice disorder (due ion migration and related point defects). Furthermore, threshold excitation density of phase segregation ($P_{th}^{Seg}$) is also calculated using different laser power densities at two different temperatures and found to be ~ 2.0 mW/cm$^2$ at 296 K and ~ 1.5 mW/cm$^2$ at 330 K (Fig. S5, supplementary information). In comparison to 296K, the observed low $P_{th}^{Seg}$ value at 330 K reveals that lower-excitation density input is required for initiating the phase segregation process at higher temperature (faster segregation rate). The estimated $P_{th}^{Seg}$ value of TC-MHP is about 100 times higher as compare to reported values of ~30 μW/cm$^2$ and ~40 μW/cm$^2$ for $MAPb(I_{0.5}Br_{0.5})_3$ at 300 K [8,24], which is meaningful as triple cation perovskites were developed for their increased robustness as compared to single cation perovskites.

**Phase segregation and carrier lifetime**

To better understand the trap-mediated recombination process and gain an insight into the charge-carrier trapping dynamics in photo-excited TC-MHP film, we performed TCSPC measurements. Fig. 2 shows the PL decay curves of red-shifted PL emission on a semi-logarithmic scale measured just after the laser excitation at different power-densities. The exponential fit of PL-decay curves exhibit two decay components as a fast ($\tau_1$) and a slow ($\tau_2$) one, as plotted in inset of Fig. 2. At lower power exposure (≤ 1 mW/cm$^2$), the calculated decay lifetime components are $\tau_1$ ~ 752 ns ± 18 ns (94%) and $\tau_2$ ~ 1985 ± 175 ns (6%), increase nonlinearly with red-shift in PL emission to maximum value of $\tau_1$ ~ 2580 ns ±78 ns (63%) and $\tau_2$ ~ 6343 ± 142 ns (37%) after increasing the excitation power-density. However, lifetime components start decreasing at higher power-densities when no more redshift occurs in PL emission peak. The average lifetime of both components is calculated by considering their weighted fraction and it is found that average lifetime increases from ~0.82 μs to ~3.94 μs as PL emission shift from 1.624 eV to 1.564 eV as a function of excitation power-density. The relative weighted contribution of $\tau_1$ component decreases from 94% to 63 % concurrently with increase in $\tau_2$ component from 6% to 37%. It confirms that during photo-induced population of low energy bandgap states, excited charge carriers in TC-MHP start funneling into these low-energy bandgap states where they live longer and release slowly.

To confirm the photo-excitation induced halide phase segregation in TC-MHP film, we have monitored its emission under continuous laser excitation at high $P_{exc}$ (840 mW/cm$^2$). We were able to record the PL emission arising from segregated bromide–rich domains even for such low initial Br-concentration (17 %) in TC-MHP film. Fig. 3 (a) shows time dependent red-shift in PL emission under continuous excitation and Fig. 3 (b) shows the evolution of a weak emission band at ~ 2.33 eV, attributed to bromide-rich domains, where PL intensity grows with exposure time and saturates within 28 min of exposure time. The segregation rate constant is calculated from time dependent exponential growth of integral intensities of bromide-and iodide-rich domains. It and found that PL emission from iodide-rich domains shows higher segregate rate ($5.58 \times 10^{-2}$ s$^{-1}$) than the bromide-rich domains ($0.93 \times 10^{-2}$ s$^{-1}$). Theoretically and experimentally it is reported that Pb-Br bond is shorter and stronger than the Pb-I bond [25,26]. Therefore, due to higher Pb-Br binding energy, bromide ions require high activation energy [27] causing slower segregation rate. Fig. 3 (c) shows the TCSPC measurement on both mixed and segregated bromide-and iodide-rich domains. PL emission decays rapidly in bromide-rich domains at 2.33 eV whereas longer PL-decay time is observed in iodide-rich domains (1.564 eV). PL decay curve from bromide-rich domains exhibits amplitude-weighted lifetimes of $\tau_1$ ~5 ns (96%) and $\tau_2$ ~24 ns (4%), respectively. The shorter carrier lifetime and significantly weaker emission from the bromide-rich domains as compared to iodide-rich region suggest that photo-generated charge carriers may quickly funneled and accumulated in the low-energy bandgap iodide-rich domains. W. Mao and coworkers reports that the low bandgap iodide-rich domains efficiently trap the free-carriers, revealing their dominant role as radiative recombination centres [28]. Similar assumptions of charge transfer between segregated phases (bromide-rich and iodide-rich domains) as well as trap-mediated recombination have been proposed in recent studies [6,13]. Therefore, it is expected that conduction/ valence band of segregated low-bandgap iodide-rich domains would serve as a sink for excited carriers from bromide-rich regions and remaining mixed TC-MHP regions, where they are long lived until recombination as schematically shown in Fig. 3 (d). The nature of trap states or trapping carrier type (electrons or holes) is not yet entirely understood in hybrid perovskite materials [29,30], although reports have suggested that electrons are more likely to be trapped than holes [14,15]. It is understood that estimated difference in the conduction band energies for $MAPbI_3$ and $MAPbBr_3$ is significantly low (~ 0.09 eV) while, the remaining difference in bandgap energy is mainly due to larger offset in their valence band energies (~200 meV) [8,27]. Therefore, valence band alignment of iodide-rich domains would act as a major energy barrier for holes to move from the iodide-rich phase back to remaining mixed phase and/or bromide-rich domains. Therefore, slow release of trapped charge carriers (mainly holes) from smaller-bandgap iodide-rich domains are likely to contribute in slower recombination and responsible for long carrier lifetime in phase segregated TC-MHP film.

**Activation energies of halide ion migration and recovery kinetics under darkness**

The recovery kinetics of phase segregated TC-MHP film is studied under darkness at different temperatures. Initially, TC-MHP film is illuminated at high excitation power-density (840 $mW/cm^2$) to achieve the phase segregated stage; then recovery process in PL emission is monitored at different time intervals using very low excitation power-density (< 1 $mW/cm^2$) as shown in supplementary Fig. S7 (a- e). Results reveal that halide phase segregation is reversible under darkness with nearly complete recovery of original PL emission spectra. In recovery process, PL emission peak recover back to original position as a function of time concurrently with decrease in PL intensity. It confirms the reduction of existing low-bandgap trapping sites of iodide rich domains occur due to remixing of halide ions under dark. The first-order phase segregation and recovery rate constants are determined from exponential fits of PL peak positions as a function of time as shown in supplementary Fig. S (6, 7). During photo-induced segregation process, the slowest segregation rate constant ($k_{seg} = 0.29 \times 10^{-3}\ s^{-1}$) is observed at lower temperature (250 K). By contrast, increasing temperature leads to fastest segregation rate as extracted values are $K_{seg} = 0.81 \times 10^{-3}\ s^{-1}$, $1.28 \times 10^{-3}\ s^{-1}$, $2.39 \times 10^{-3}\ s^{-1}$, $8.92 \times 10^{-3}\ s^{-1}$ and $20.29 \times 10^{-3}\ s^{-1}$ for 260 K, 270 K, 280 K, 296 K and 310 K, respectively. Similar trends in recovery rate constants are observed at different temperatures under darkness. In Fig. 4 (a), results confirm that recovery dynamics is highly temperature sensitive. However, the overall recovery process in dark occurs on a much slower timescale than the photo-induced phase segregation process at given temperature. At room temperature (296 K), segregated TC-MHP film recovers itself completely in 10 h under dark. Moreover, increase in temperature leads to faster recovery of segregated films as ~5 h at 310 K, 2.5 h for 320 K and less than 2 h at 330 K, respectively. Similarly, corresponding recovery rate constant ($k_{rec}$) is dramatically increased from $1.32 \times 10^{-4}$ $s^{-1}$ to $6.32 \times 10^{-4}$ $s^{-1}$, as film heating temperature raised from 296 K to 330 K. However, decrease in temperature to below 270 K, the recovery kinetics is almost frozen since the PL emission recovers in ~ 24 h at 270 K with an very slow rate ($k_{rec} = 0.54 \times 10^{-4}$ $s^{-1}$). Thus, decreasing in temperature leads to reduction of halide ion mobility in TC-MHP film. Similar results are reported by our co-workers that ionic conduction is frozen below 263 K in $MAPI_3$ perovskite based solar cells [31].

The activation energy of halide ion movement is estimated in two different photo-excitation induced phase -segregation ($E_a^{Seg.}$) and dark-recovery ($E_a^{Rec.}$) processes. In Fig. 4 (b), an Arrhenius plot is constructed from the natural log of the rate constants (k) obtained at different temperatures versus 1000/T, using the following Arrhenius equation:

$$\ln(k) = -\frac{E_a}{RT} + \ln(A) \tag{1}$$

where $E_a$ is the activation energy, $R$ is the universal gas constant and $A$ is a pre-exponential factor. Activation energies of halide ion movement in two different processes are calculated by least square fitting of linear data points and reveals that activation energy in phase segregation process $E_a^{Seg.}$ is 0.46 ± 0.02 eV which is found to be larger than in case of recovery process, $E_a^{Rec.}$ ~ 0.31 ± 0.03 eV. Indeed, calculated $E_a$ values are consistent with the reported values of forward photo-segregation activation energies of order ~0.27–0.28 eV for $MAPb(I_{0.5}Br_{0.5})_3$ [6,24]. Moreover, the calculated $E_a$ values are also consistent with prior studies of halide ion motion, where activation energies of halide vacancy $Br^-$ ($I^-$) have been reported in between 0.09 and 0.27 eV (0.08 and 0.58 eV), respectively [6,32–34]. The differences in reported activation energy may in part be due to different local stoichiometry of perovskite structure (different halide concentration and incorporation of different cations) which influences the halide ion mobility. In present case, two different $E_a$ values confirm the two different halide ion migration mechanisms of both phase segregation and recovery processes. The high $E_a^{Seg.}$ value in phase segregation process indicates to high activation energy is required to alter the lead-halide bonds for initiating halide ion migration. On the other hand, comparably low $E_a^{Rec.}$ value in recovery process may be due to the fact that back-diffusion of halide ions is driven by halide concentration gradient [34] or entropically driven intermixing [35,36] to return the perovskite into a homogeneous condition. In other possible way, once the trapped charge carriers have recombined, the created local electric field vanishes and then the halide ions start back-diffusion to fill the vacancies left in the original position.

## Low-temperature dynamics and role of exciton-phonon coupling

The temperature dependent PL measurements are performed for in-depth understanding of mixed and demixed stage (segregated phases) in TC-MHP film over a range of temperatures ($T$ = 300–10 K). Fig. 5 (a) shows the temperature-dependent PL spectra of mixed stage (parent phase), on a semi-logarithmic scale measured at low excitation power-density (< 1 mW/cm$^2$) and exhibits a single emission peak with low energy asymmetry at low-temperature (10 K). The low-energy asymmetry in PL emission is related to presence of bound excitons and shallow energy levels by iodine interstitials defects [37,38]. Fig 5 (b) shows the semi-log scale temperature dependent PL spectra of demixed stage measured at high power-density (840 mW/cm$^2$). For demixed stage dynamics, first sample was illuminated at high power (840 mW/cm$^2$) at room temperature to create phase segregated domains and then cooled down to 10 K under continuous photo-excitation. In Fig. 5 (b) results shows the presence of two emission peaks located at 1.574 and 1.534 eV at 10 K, where, low-energy emission peak (LE-peak) is related to iodide-rich phase and high-energy emission peak (HE-peak) corresponds to remaining mixed halide phase. Fig. 5 (c) shows the blue-shift response of PL emission in both mixed and demixed stages with increasing temperature. In demixed stage, both peaks are red-shifted as compared to emission peak of mixed stage caused due to local change in halide composition of TC-MHP film after phase segregation. Furthermore, irrespective of perovskite composition, PL emission peaks of both mixed and demixed stages exhibit non-liner blue-shift when temperature rises from 10 K to 300 K. Generally, this behavior of perovskites is related to the thermal expansion of the lattice and stabilization of valence band energy [39]. The response character of blue-shift with temperature changes near T ~ 150 K in both stages, revealing the standard phase transition from orthorhombic to tetragonal phase in TC-MHP film. In mixed stage, PL emission peak experiences a blue-shift of 18 meV as rising the temperature from 10 K to 80 K, while after phase transition to tetragonal phase, the PL peak exhibits a blue-shift of 24 meV when rising temperature from 150 K to 300 K. In demixed stage, HE-peak follows the similar response and exhibits the blue-shift of ~15 meV and ~25 meV in orthorhombic (10 K - 80 K) and tetragonal phase (150 K - 300 K), respectively. However, LE-peak (iodide-rich domain) in demixed stage shows blue-shift of ~18 meV in orthorhombic phase and large blue-shift of ~ 45 meV in tetragonal phase when temperature goes from 150 K to 300 K.

Furthermore, we carefully examine the temperature dependent PL linewidth broadening to study the contributions of charge-carrier interactions with phonons in both mixed and demixed stages using Segall's expression [40].

$$\Gamma = \Gamma_{inh} + \Gamma_{ac} + \Gamma_{LO} + \Gamma_{imp} \quad (2)$$

$$\Gamma = \Gamma_{inh} + \gamma_{ac} T + \frac{\gamma_{LO}}{exp(\hbar\omega_{LO}/k_B T) - 1} \quad (3)$$

where $\Gamma_{inh}$ is the inhomogeneous broadening contribution that arises from scattering due to exciton–exciton interactions and crystal disorder, and is temperature independent. In expression, $\Gamma_{ac}$ and $\Gamma_{LO}$ are the homogeneous broadening terms resulting from acoustic and longitudinal optical (LO)-phonon (Fröhlich) scattering with charge carrier-phonon coupling strength of $\gamma_{ac}$ and $\gamma_{LO}$, respectively. Acoustic phonons, whose energy is much smaller than $k_B T$, mainly related to the deformation potential interaction which is linearly dependent on temperature. The exciton LO-phonon coupling (Fröhlich coupling) coefficient is associated with Bose–Einstein distribution of the LO-phonons, where $\hbar\omega_{LO}$ is the optical phonon energy of weakly dispersive LO-phonon branch. The FWHM of emission peaks are plotted as a function of temperature along with the least square fitting using Equation (3). The inset of Fig. 5 (d) demonstrates the typical PL linewidth broadening pattern with temperature associated with different scattering mechanisms. Result makes it apparent that Fröhlich coupling to LO-phonons is the predominant cause of linewidth broadening at higher temperature. Therefore, the scattering term from ionized impurities ($\Gamma_{imp}$) is excluded in Equation (3) for rest of the analysis. In mixed stage, the calculated exciton LO-phonon coupling strength is ~ 14.82 meV with LO-phonon energy of ~ 22.8 meV. In demixed stage, HE-peak follows a similar response of linewith broadening with increasing temperature; the corresponding coupling strength and LO-phonon energy are ~ 38.86 meV and 86 meV, respectively. On the other hand, LE-peak shows large broadening in linewidth with temperature and the corresponding Fröhlich coupling strength and LO-phonon energy are ~ 46.70 meV and ~168 meV, respectively. Therefore strong exciton LO-phonon coupling is observed in LE-peak of demixed stage, which may be due to the strong interaction of phonon with long lived charge carriers trapped in low energy iodide-rich domains.

## Discussion

In the different scenario of photo-excitation induced phase segregation in mixed halide perovskite, excited-charge carriers play a main role either caused by light illumination [41,42] or by electrical injection [7,43–46]. In present case, photo-excited phase segregation would be initiate by charge separation arising from the carrier trapping on surface defects sites and at the grain boundaries (e.g., electrons and leaving behind holes) which attribute to the formation of local electric fields [34,47,48]. Furthermore, ion migration induced segregated domains having a significant offset in valence band and conduction band energies create a degree of charge separation between the randomly distributed segregated domains and remaining mixed halide phase, which thus establishes a local electric field in the film. The high barrier in valence-band energies between both iodide-rich domains and mixed halide perovskite act to draw holes away from the electrons. Thus photo-excitation induced local electric field promotes the degree of halide segregation and charge separation by further trapping of excited carriers into smaller-energy bandgap iodide-rich domains in a feedback loop. The phase segregated halide domains are stabilized under photo-excitation when a steady state is reached. A high Fröhlich coupling is observed in these iodide-rich domains due to strong interaction of soft ionic lattice with long lived excited charge carriers. It is reported that segregated halide clusters are stabilized by the presence of a photo-generated trapped polaron in the hybrid perovskites [18]**.** Therefore, high exciton LO-phonon coupling in iodide-rich domains induce sufficient polaronic strain which is able to locally change the free energy for halide demixing and leads to stabilize the segregated domains under photo-excitation. Eventually, concentration gradient of halide ions and limited funneling of photo-excited charge carriers into iodide-rich domains are also limits the further growth of segregated domains under photo-excitation.

## Conclusions

In summary, our findings provide new insights into the halide ion migration and phase segregation effect in triple-cation based mixed-halide perovskite film when subjected to photo-excitation. We demonstrate that laser-excitation induced phase segregation in TC-MHP film leads to formation of smaller-bandgap iodide-rich and larger-bandgap bromide-rich domains where iodide-rich domains efficiently traps the photo-excited-carriers, revealing their dominant role in the origin of unusual long carrier lifetime (larger than 1 μs). Moreover, we found that phase segregation process is fully reversible under dark and recovery rate increases with rising temperature. It is also shown that activation energy of halide ion migration is higher during photo-excitation induced phase segregation process as compared to recovery process under darkness. In addition, temperature-dependent PL studies have been performed for better understanding of exciton–phonon coupling in both mixed and demixed (segregated) states. A high Fröhlich coupling is observed in the segregated iodide-rich domains as compare to demixed (parent) phase due to strong coupling of phonon with long lived trapped charge carriers. Thus high coupling strength in segregated iodide-rich domains may be responsible for the limited growth and stabilization of segregated halide domains under photo-excitation. These findings will help to understand the key issues of phase segregation in the mixed halide perovskite materials for the development of efficient solar cells and optoelectronic devices.

## Acknowledgements

This work was supported by the LabEx PALM (ANR-10-LABX-0039-PALM), and IRS MOMENTOM (Université Paris-Saclay).

# Figure captions

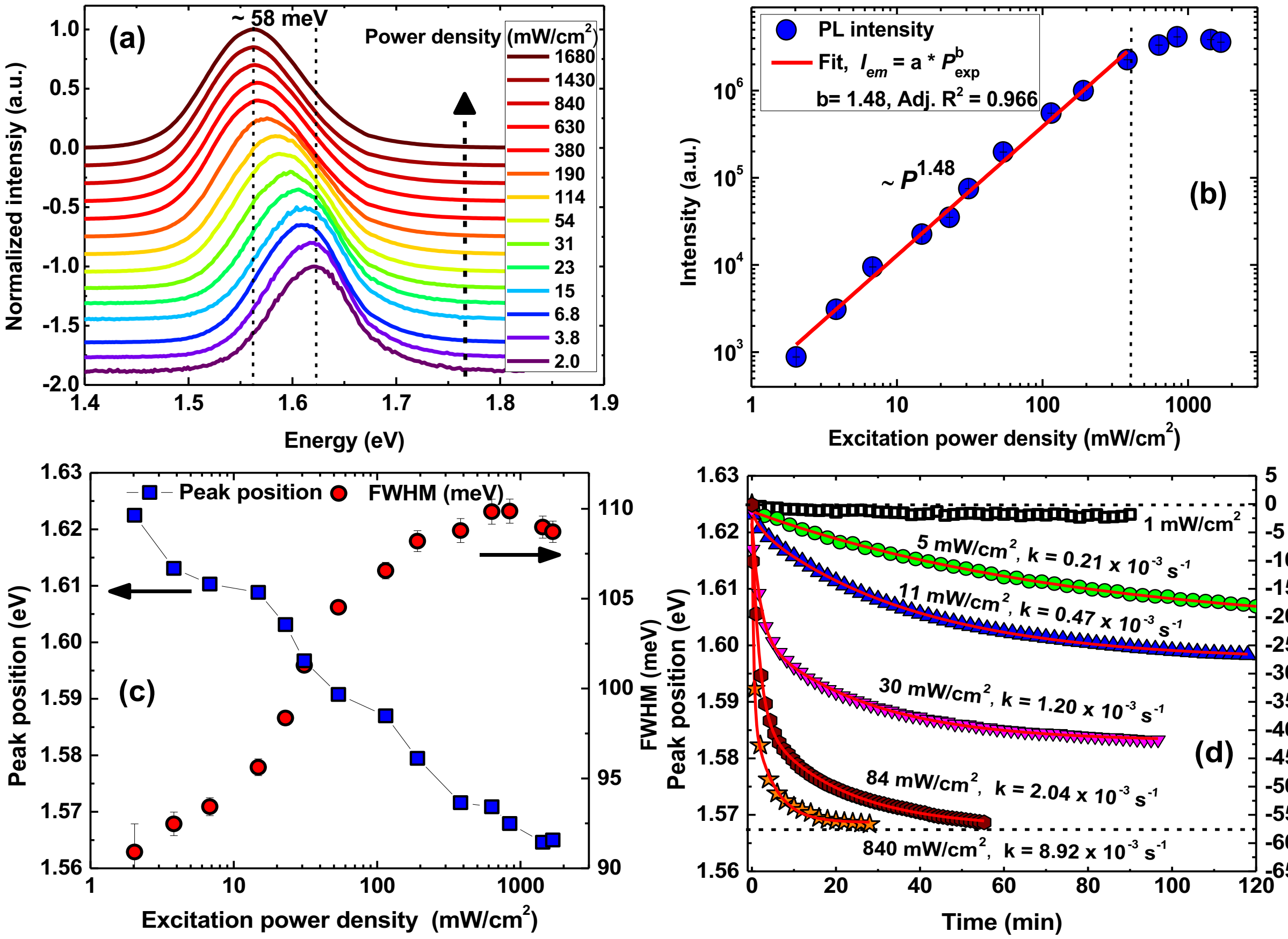

**Figure 1: (a)** Power-dependent photoluminescence spectra of triple-cation mixed-halide $(MA_{0.17}\,FA_{0.83})_{0.95}Cs_{0.05}\,Pb(I_{0.83}\,Br_{0.17})_3$ perovskite film showing red shift in photoluminescence with increasing excitation power-density at 296 K. **(b)** Logarithmic plot, showing increase in integrated PL intensity as a power law of excitation power-density, with an exponent of ~3/2 (red line). **(c)** Plot of red shift in PL emission peak and line-width broadening as a function of laser power density. **(d)** Time dependent red-shift in PL emission under different excitation power-densities and fitted exponential curves to calculate phase segregation rate constants.

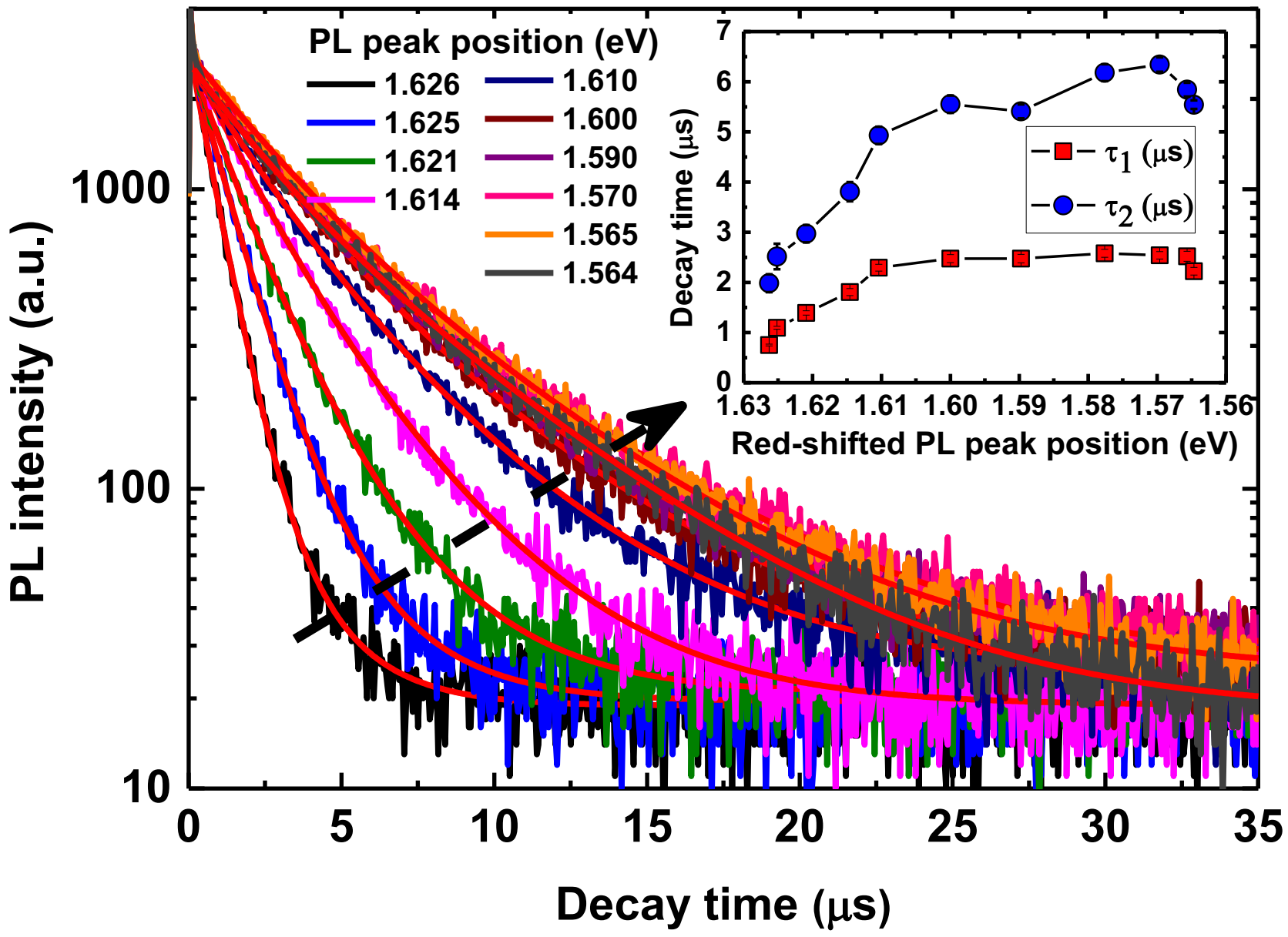

**Figure 2:** PL decay spectra measured on red-shifted PL emission position after light soaking of different excitation power-densities at room temperature (296 K) with fitted exponential decay curves. Inset: Lifetime components as a function of PL emission position.

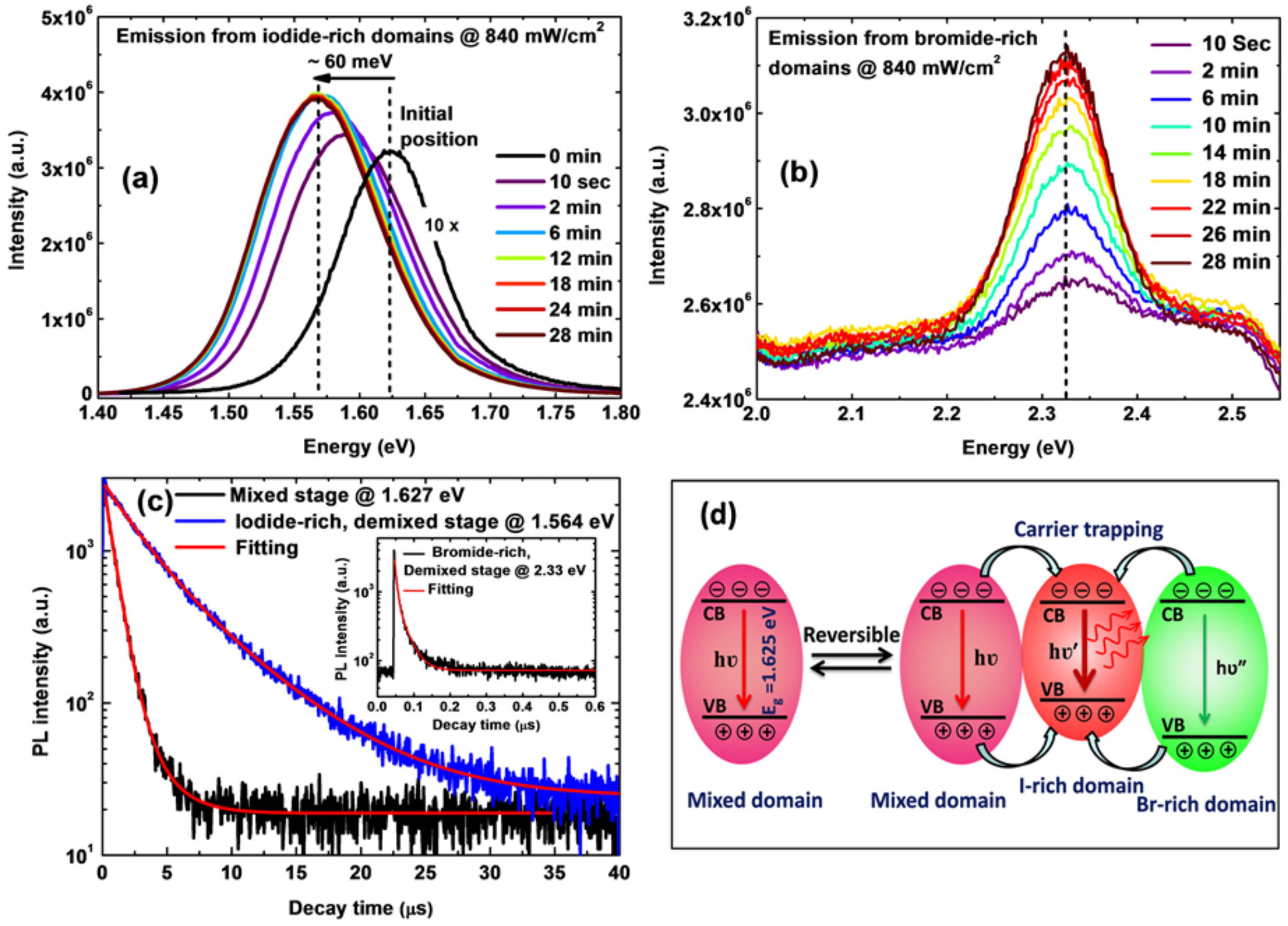

**Figure 3:** Spectroscopic observation of halide phase segregation in triple-cation mixed-halide $(MA_{0.17}\ FA_{0.83})_{0.95}Cs_{0.05}\ Pb(I_{0.83}\ Br_{0.17})_3$ perovskite film. **(a)** Time dependent red-shift in PL emission and evolution of PL spectra from iodide-rich domains under continuous wave excitation ($P_{exc}$ = 840 mW/cm$^2$). **(b)** Time dependent evolution of PL spectra from bromide-rich domains under continuous wave excitation ($P_{exc}$ = 840 mW/cm$^2$). **(c)** A comparative analysis of PL decay spectra from initial mixed stage and segregated iodide and bromide-rich domains. Inset: PL decay spectra from bromide-rich domains shows faster decay time. **(d)** Schematic illustrating the reversible phase segregation process where carrier trapping and recombination in segregated smaller bandgap iodide-rich domain generates red-shifted PL emission.

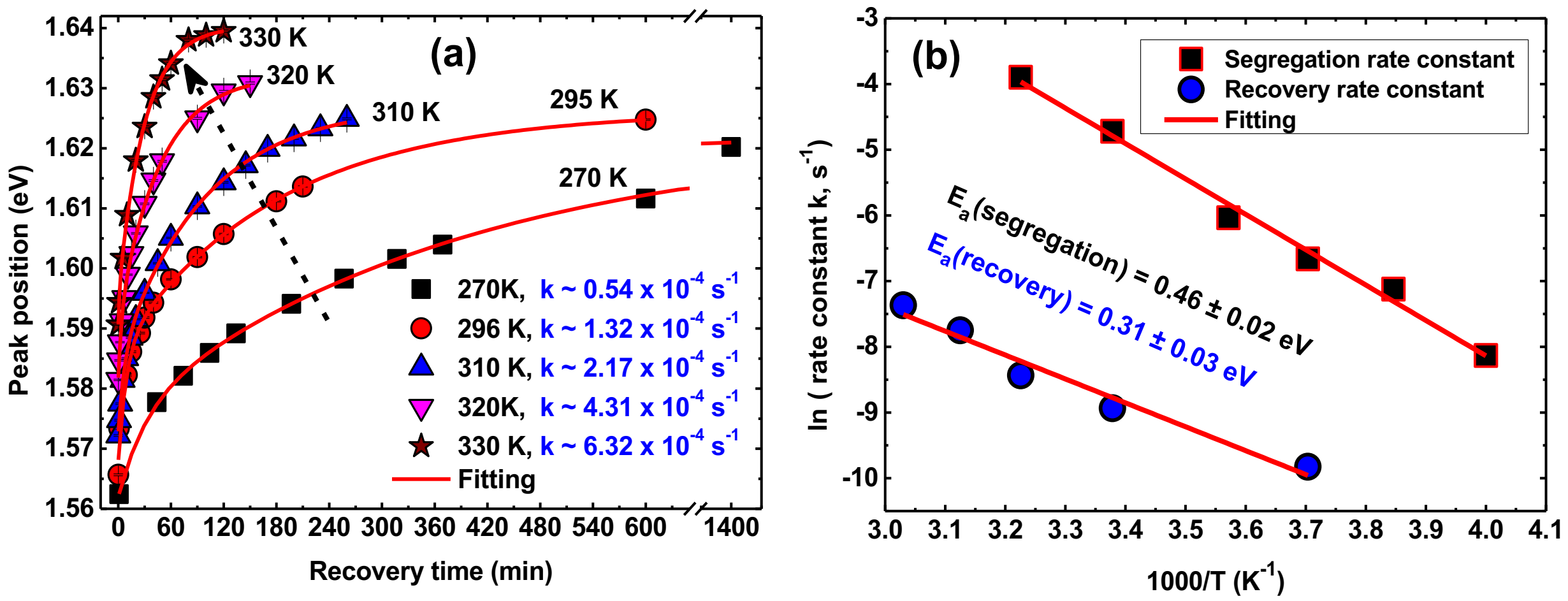

**Figure 4: (a)** Time dependent recovery rate dynamics of phase segregated triple-cation mixed-halide $(MA_{0.17}\,FA_{0.83})_{0.95}Cs_{0.05}\,Pb(I_{0.83}\,Br_{0.17})_3$ perovskite film under dark at different temperatures with exponential fitted curves. **(b)** Arrenhius plots of ln (segregation and recovery rate constants) versus 1000/T for activation energies of halide ion migration during phase segregation and recovery processes.

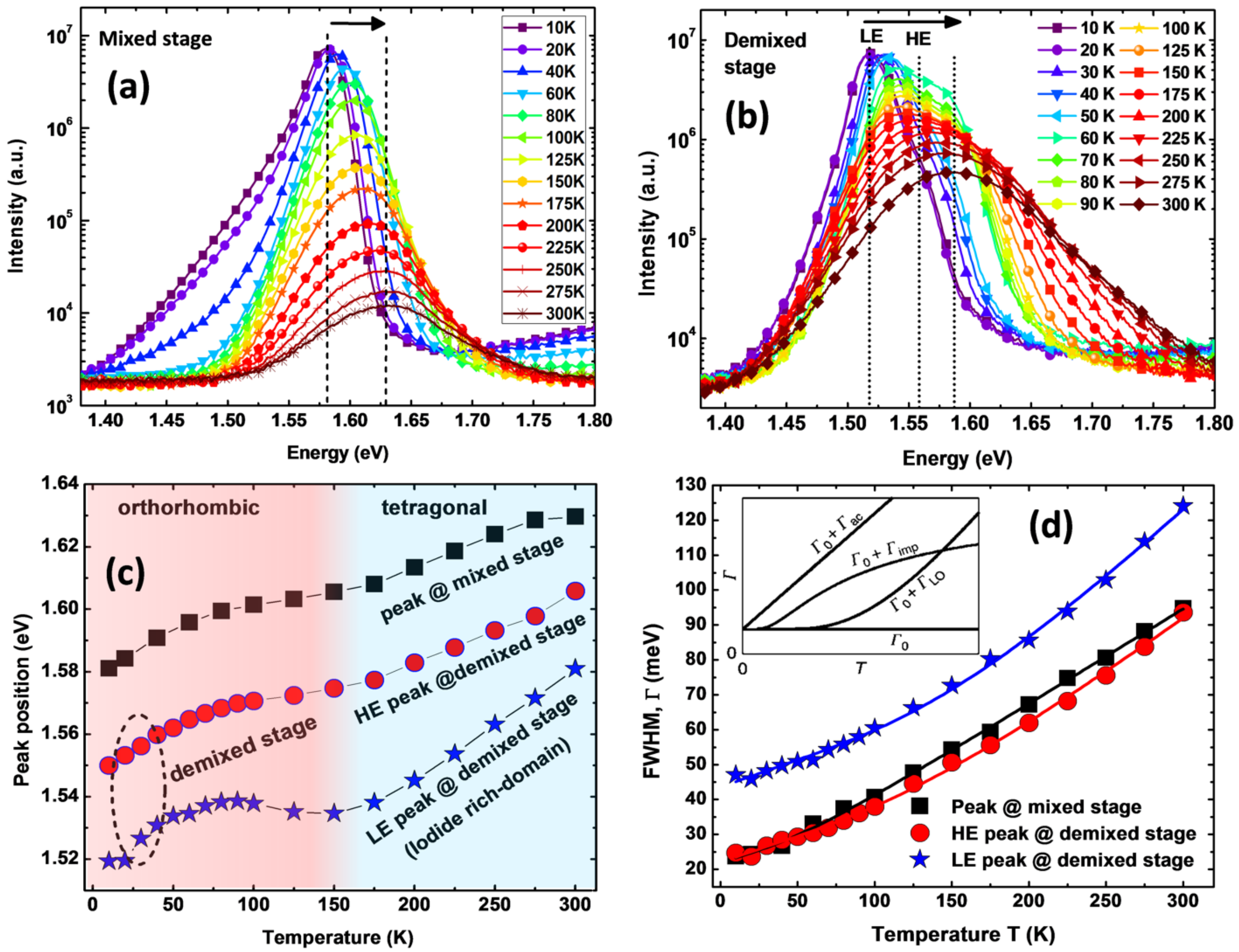

**Figure 5:** Temperature dependent PL emission characteristics of triple-cation mixed-halide $(MA_{0.17}\ FA_{0.83})_{0.95}Cs_{0.05}\ Pb(I_{0.83}\ Br_{0.17})_3$ perovskite film in both mixed and demixed (segregated phase) stages. **(a)** Temperature-dependent PL spectra at low excitation power-density ($P_{exc}$ = 1 mW/cm$^2$) and **(b)** PL spectra of demixed stage under excitation with high power-density ($P_{exc}$ = 840 mW/cm$^2$). **(c)** A comparison of blue-shift response of emission peaks in both mixed and demixed stages as a function of temperature. **(d)** The linewidth broadening of PL spectra in both mixed and demixed stages as function of temperature (symbols) along with least square fitting curves (solid lines) to examine the exciton- phonon coupling.